\documentclass[12pt]{article}
\usepackage{amsmath, amssymb}
\usepackage[latin1]{inputenc}
\usepackage{mathrsfs}
\usepackage{cite}
\usepackage{dsfont}

\begin{document}
\date{}
\title{Algebraic approach and coherent states for a relativistic quantum particle in cosmic string spacetime}

\author{M. Salazar-Ram\'{\i}rez$^{a}$\footnote{{\it E-mail address:} escomphysics@gmail.com.mx}, D. Ojeda-Guill\'en$^{a}$ and\\ R. D. Mota$^{b}$ } \maketitle

\begin{minipage}{0.9\textwidth}

\small $^{a}$ Escuela Superior de C\'omputo, Instituto Polit\'ecnico Nacional,
Av. Juan de Dios B\'atiz esq. Av. Miguel Oth\'on de Mendiz\'abal, Col. Lindavista,
Del. Gustavo A. Madero, C.P. 07738, Ciudad de M\'exico, Mexico.\\

\small $^{b}$ Escuela Superior de Ingenier{\'i}a Mec\'anica y El\'ectrica, Unidad Culhuac\'an,
Instituto Polit\'ecnico Nacional, Av. Santa Ana No. 1000, Col. San
Francisco Culhuac\'an, Del. Coyoac\'an, C.P. 04430, Ciudad de M\'exico, Mexico.\\

\end{minipage}

\begin{abstract}
We study a relativistic quantum particle in cosmic string spacetime in the presence
of a uniform magnetic field and a Coulomb-type scalar potential. It is shown that the radial part of this problem possesses the
$su(1,1)$ symmetry. We obtain the energy spectrum and eigenfunctions of this problem by using two algebraic methods:
the Schr\"odinger factorization and the tilting transformation. Finally, we give the explicit form of the relativistic coherent
states for this problem.
\end{abstract}

\section{Introduction}

The cosmic strings, which are hypothetical $1$-dimensional topological defects were introduced in the $1970s$ \cite{TKIB}. In the $1980s$ and the early $90s$ a strong interest arose in connection with cosmic strings \cite{JPOL}. This interest is due to the bridge that these cosmic strings provide between the physics of the very small and the very large. It is known from the literature \cite{MBH,VILEN} that cosmic strings are predicted in some unified theories of particle interactions and could have formed in one of the numerous phase transitions in the early universe due to the Kibble mechanism. These cosmic strings could also be responsible for the large-scale structure of the universe.

The study of quantum systems under the influence of the gravitational field have been of great interest in particle physics. A problem that has been extensively studied is the appearance of topological phases in the quantum dynamics of a single particle moving freely in multiple connected space-times. As an example of a gravitational effect of topological origin we can consider the analogue of the electromagnetic Aharonov-Bohm effect \cite{FORD,BEZ}. This effect is provided when a particle is transported around an idealized cosmic string \cite{VILEN1,JRG,WAH,LINET} along a closed curve in which case the string is noticed at all. The influence of the Aharonov-Casher effect \cite{YAHA} has been recently studied in the Minkowski spacetime, the cosmic string spacetime and the cosmic dislocation spacetime \cite{KBAK}. In reference \cite{KBAK} the authors solved the Dirac equation and obtained the energy levels for bound states and the Dirac spinors for positive-energy solutions. Recently, the relativistic quantum motion of charged spin-$0$ and spin-$\frac{1}{2}$ particles in the presence of a uniform magnetic field and scalar potentials in the cosmic string spacetime has been studied \cite{MAD}. The Dirac equation of this problem was treated in reference \cite{YESI} in an algebraic way.

On the other hand, the factorization method introduced by Infeld and Hull \cite{INF1,INF2} has been of great importance in the study of quantum systems. In their work, Infeld and Hull gave a systematic method to factorize and classify a large class of potentials. However, Dirac \cite{DIRLIB} and Schr\"odinger \cite{SCH1A,SCH1B,SCH1C} established the fundamental ideas of factorization in quantum physics. The importance of these factorization methods lies in the fact that if the Schr\"odinger equation is factorizable, the energy spectrum and eigenfunctions are obtained in an algebraic way. Moreover, the operators constructed from these methods are related to compact and non-compact Lie algebras.

Besides, the existence of a symmetry group has been used to compute the coherent states of many physical problems \cite{GAZL,KLIL,KLAUL,PER,ZHANG}. The harmonic oscillator coherent states were introduced by Schr\"odinger while he was looking for a system which possessed a classical behavior \cite{SCH}. The importance of coherent states in quantum optics was studied in references \cite{GLAU,KLAU,KLAU2,SUDAR} and the generalization of these states to any algebra of a symmetry group is reported in references \cite{BANDG,PEREL,AREC}.

The aim of the present work is to study a relativistic quantum particle in cosmic string spacetime in the presence of a uniform magnetic field and scalar potential. In order to obtain the energy spectrum and the eigenfunctions of this problem we use the theory of unitary representations and two algebraic methods: The Schr\"odinger factorization method and the tilting transformation. From the $SU(1,1)$ group theory we construct the relativistic coherent states for the radial part.

This work is organized as follows. In Section $2$ we obtain the uncoupled second-order differential equations satisfied by the radial components. In Section $3$, we apply the Schr\"odinger factorization method to one of the uncoupled equations to obtain the energy spectrum and the eigenfunctions of our problem. In Section $4$, the energy spectrum and the eigenfunctions are obtained by using the tilting transformation and a realization of the $su(1,1)$ Lie algebra which is energy-independent. In Section $5$ we obtain the explicit expression of $SU(1,1)$ Perelomov coherent states for a relativistic quantum particle in cosmic string spacetime in the presence
of a uniform magnetic field and scalar potential. Finally, we give some concluding remarks.

\section{The Dirac equation for an arbitrary curved spacetime}

The metric tensor for the cosmic string spacetime in cylindrical coordinates is defined by the line element
 \begin{equation}
ds^2=dt^2-dr^2-\rho^2 r^2d\phi^2-dz^2,
\end{equation}
where the coordinates $(t,z)\in(-\infty,\infty)$, $r\geq 0$, the angular variable $\phi\in[0,2\pi]$ and $\rho=1-4\mu$ is related to the deficit angle, with $\mu$ the linear mass density. Let's consider
a uniform magnetic field parallel to the string $\overrightarrow{B}=\overrightarrow{\nabla}\times \overrightarrow{A}=B_0 \hat{k}$. Thus, in the Coulomb gauge the vector potential can be expressed by
\begin{equation}
\overrightarrow{A}=(0,A_{\phi},0), \quad\quad A_{\phi}=-\frac{1}{2}\rho B_0 r^2.
\end{equation}

The Dirac equation to an arbitrary curved spacetime, in the presence of an external electromagnetic potential and a scalar potential can be written as
\begin{equation}\label{EDir}
\left[i\gamma^{\mu}(x)\left(\nabla_{\mu}+ieA_{\mu}\right)-\left(M+S(r)\right)\right]\Psi(x)=0,
\end{equation}
with $M$ the mass of the particle and $\gamma^{\mu}$ the generalized Dirac matrices, which satisfy the Clifford algebra $\left\{\gamma^{\mu}(x),\gamma^{\nu}(x)\right\}=2g^{\mu\nu}$. These generalized Dirac
matrices are defined in terms of a set of tetrad fields and the constant Dirac matrices as
\begin{equation}\label{tetra1}
\gamma^{\mu}(x)=e_{(a)}^{\mu}(x)\gamma(a).
\end{equation}
The covariant derivative for fermions field is given by
\begin{equation}\label{EDir2}
\nabla_{\mu}=\partial_{\mu}+\Gamma_{\mu},
\end{equation}
where $\Gamma_{\mu}$ is the spinor affine connection
\begin{equation}\label{spincon}
\Gamma_{\mu}=\frac{1}{4}\gamma^{(a)}\gamma^{(b)}e_{(a)}^{\nu}\left[\partial_{\mu}e_{(b)\nu}-\Gamma_{\mu\nu}^\sigma e_{(b)\sigma}\right].
\end{equation}
Here, $\Gamma_{\mu\nu}^\sigma$ is the Christoffel symbol of the second kind and the tetrad basis $e_{(a)}^{\mu}(x)$ satisfies the relation \cite{MAD}
\begin{equation}\label{EDir3}
\eta^{(a)(b)}e_{(a)}^{\mu}(x)e_{(b)}^{\nu}(x)=g^{\mu\nu}.
\end{equation}
The matrices $\gamma^{(a)}$ are the standard flat spacetime Dirac matrices defined as
\begin{equation}\label{matr1a}
\gamma^{0}=\begin{pmatrix}
\textbf{1} & 0 \\
0 & \textbf{-1}
\end{pmatrix}, \hspace{0.5cm} \gamma^{a}=\begin{pmatrix}
0 & \sigma^{(a)} \\
-\sigma^{(a)} & 0
\end{pmatrix},
\end{equation}
where $\sigma^{(a)}$ are the Pauli matrices. The tetrad basis $e_{(a)}^{\mu}(x)$ in the spacetime of a cosmic string is written as \cite{MAD}
\begin{equation}\label{matte}
e_{(a)}^{\mu}=\begin{pmatrix}
1 & 0 & 0 & 0 \\
0 & \cos{\phi} & -\frac{\sin{\phi}}{\rho{r}} & 0\\
0 & \sin{\phi} & \frac{\cos{\phi}}{\rho{r}} & 0\\
0 & 0 & 0 & 1
\end{pmatrix}.
\end{equation}
Therefore, from equations (\ref{tetra1}), (\ref{matr1a}) and (\ref{matte}) we obtain the explicit form of the generalized Dirac matrices
\begin{equation}\label{matr1}
\gamma^{0}=\gamma^{(0)},\hspace{0.5cm}\gamma^{i}=\begin{pmatrix}
0 & \sigma^{i} \\
-\sigma^{i} & 0
\end{pmatrix}.
\end{equation}
In this expression, $\sigma^{\mu}$ represents the modified Pauli matrices, which are given by
\begin{equation}\label{matr2}
\sigma^{1}=\sigma^{r}=\begin{pmatrix}
0 & e^{-i\phi} \\
e^{i\phi} & 0
\end{pmatrix}, \hspace{0.5cm} \sigma^{2}=\sigma^{\phi}=-\frac{i}{\rho{r}}\begin{pmatrix}
0 & e^{-i\phi} \\
-e^{-i\phi} & 0
\end{pmatrix},
\end{equation}
and
\begin{equation}
\sigma^{3}=\sigma^{z}=\sigma^{(3)},\hspace{0.5cm}\text{hence}\hspace{0.5cm}\gamma^{3}=\gamma^{(3)}.
\end{equation}
From equations (\ref{spincon}), (\ref{matte}) and (\ref{matr1}) we obtain the nonzero components $\Gamma_{\phi}$ of the spin connection
\begin{equation}
\Gamma_{\mu}=\left(0,0,\Gamma_{\phi},0\right),\hspace{0.5cm}\text{with}\hspace{0.5cm} \Gamma_{\phi}=\frac{i\left(1-\rho\right)}{2}\Sigma^{3},
\end{equation}
where
\begin{equation}
\Sigma^{3}=\begin{pmatrix}
\sigma^{(3)} & 0 \\
0 & \sigma^{(3)}
\end{pmatrix}.
\end{equation}
With these previous results we can write the Hamiltonian of equation (\ref{EDir}) as
\begin{equation}\label{ham}
H=-i\gamma^{(0)}\left[\gamma^{r}\partial_{r}+\gamma^{\phi}\left(\partial_{\phi}+\frac{i\left(1-\rho\right)}{2}\Sigma^3+ieA_{\phi}(r)\right)+\gamma^{(3)}\partial_z+i\left(M+S(r)\right)\right].
\end{equation}
This Hamiltonian satisfies the commutation relations for the momentum operator and the total angular momentum operator in the $z$-direction  $\left[H,\widehat{p}_3\right]=\left[H,\widehat{J}_3\right]=0$, with $\widehat{p}_3=-i\partial_z$, $\widehat{J}_3=\widehat{L}_3+\widehat{S}_3=-i\partial_{\phi}+1/2\Sigma^{3}$. Hence, these operators satisfy the eigenvalue equations
\begin{equation}\label{ecuw}
H\Psi=E\Psi,\hspace{0.5cm}\widehat{p}_3\Psi=k\Psi, \hspace{0.5cm} \widehat{J}_3\Psi=-i\partial_{\phi}\Psi+\frac{1}{2}\Sigma^{3}\Psi=j\Psi,
\end{equation}
where $j=m+1/2$, $m=0\pm1\pm2,...$ and $k\in\left(-\infty,\infty\right)$.
We propose the Dirac wave function of equation (\ref{ecuw}) to be of the form
\begin{equation}\label{anz}
\Psi(t,r,\phi,z)=\frac{1}{\sqrt{r}}e^{(-iEt+im\phi+ikz)}
\begin{pmatrix}
F_{+}(r)\\
-iF_{-}(r)e^{i\phi}\\
G_{+}(r)\\
iG_{-}(r)e^{i\phi}
\end{pmatrix}.
\end{equation}

Thus, by using equations (\ref{ham}) and (\ref{anz}) and the discrete symmetry $G_{+}=\lambda{F_{+}}$ and $F_{-}=\lambda{G_{-}}$, we obtain the coupled differential equations
\begin{align}\label{Ed1}
\frac{dG_{-}}{dr}&+\left(\frac{j}{\rho{r}}-M\omega\right)G_{-}+k\lambda{F_{+}}+\left(M-E+S(r)\right)F_{+}=0,\\\label{Ed2}
\frac{dG_{+}}{dr}&-\left(\frac{j}{\rho{r}}-M\omega\right)G_{+}+\frac{k}{\lambda}{F_{-}}+\left(M-E+S(r)\right)F_{-}=0,\\\label{Ed3}
\frac{dF_{-}}{dr}&+\left(\frac{j}{\rho{r}}-M\omega\right)F_{-}-\frac{k}{\lambda}{G_{+}}+\left(M+E+S(r)\right)G_{+}=0,\\\label{Ed4}
\frac{dF_{+}}{dr}&-\left(\frac{j}{\rho{r}}-M\omega\right)F_{+}-k\lambda{G_{-}}+\left(M+E+S(r)\right)G_{-}=0,
\end{align}
where $\omega=\frac{eA_0}{2M}$, $\lambda=\frac{E+s\sqrt{E^2-k^2}}{k}$, $s=\pm1$ and $S(r)=\frac{s_1}{r}+s_2$. Therefore, the equations (\ref{Ed1}) and (\ref{Ed4}) can be written in the following matricial form
\begin{equation}\label{dg}
\begin{pmatrix}
\frac{dF_{+}}{dr}\\
\frac{dG_{-}}{dr}
\end{pmatrix}+
\frac{1}{r}\begin{pmatrix}
-\frac{j}{\rho} & s_1 \\
s_1 & \frac{j}{\rho} \end{pmatrix}
\begin{pmatrix}
F_{+}\\
G_{-}
\end{pmatrix}=\begin{pmatrix}
-M\omega & -\left(M+E-k\lambda+s_2\right)\\
-\left(M-E+k\lambda+s_2\right) & M\omega\end{pmatrix}\begin{pmatrix}
F_{+}\\
G_{-}
\end{pmatrix}.
\end{equation}

A similar equation holds for the other two spinor components $F_-$ and $G_+$. In order to diagonalize the $1/r$ term, we use a transformation introduced by Thaller \cite{THALL1} and Sukumar \cite{SUKU}, which implies to define a matrix $\mathbb{M}$ that satisfies the property
\begin{equation}
\mathbb{M}^{-1}\begin{pmatrix}
-\frac{j}{\rho} & s_1 \\
s_1 & \frac{j}{\rho}
\end{pmatrix}\mathbb{M}=\begin{pmatrix}
\gamma & 0 \\
0 & -\gamma
\end{pmatrix},
\end{equation}
where $\gamma=\frac{\sqrt{j^2+\rho^2s_1^2}}{\rho}$. With this requirement we obtain that the matrix $\mathbb{M}$ is given by
\begin{equation}\label{matM}
\mathbb{M}=\begin{pmatrix}
\gamma-\frac{j}{\rho} & -s_1 \\
s_1 & \gamma-\frac{j}{\rho}
\end{pmatrix}.
\end{equation}
Thus, equation (\ref{dg}) can be written as
\begin{equation}\label{mard1}
\begin{pmatrix}
\frac{d}{dr}+\frac{\gamma}{r}-\frac{M}{\rho\gamma}\left(\omega j-\rho s_1-\frac{\rho s_1 s_2}{m}\right) &-\sqrt{E^2-k^2}-\frac{M\omega\rho s_1+\left(M+s_2\right)j}{\gamma\rho} \\
-\sqrt{E^2-k^2}+\frac{M\omega\rho s_1+\left(M+s_2\right)j}{\gamma\rho} &-\frac{d}{dr}+\frac{\gamma}{r}-\frac{M}{\rho\gamma}\left(\omega j-\rho s_1-\frac{\rho s_1 s_2}{m}\right)
\end{pmatrix}\begin{pmatrix}
F\\
G
\end{pmatrix}=0,
\end{equation}
where we have introduced the definition
\begin{equation}\label{minver}
\begin{pmatrix}
F\\
G
\end{pmatrix}= \mathbb{M}^{-1}\begin{pmatrix}
F_{+}\\
G_{-}
\end{pmatrix}= \begin{pmatrix}
\frac{1}{2\gamma}&\frac{\rho s_1}{2\gamma(\gamma\rho-j)}\\
-\frac{\rho s_1}{2\gamma(\gamma\rho-j)}&
\frac{1}{2\gamma}\end{pmatrix}\begin{pmatrix}
F_{+}\\
G_{-}
\end{pmatrix}.
\end{equation}
The matrix $\mathbb{M}$ of equation (\ref{mard1}) decouple the differential equations for $F_+$ and $G_-$ as follows
\begin{equation}\label{mard2}
\begin{pmatrix}
-\frac{d^2}{dr^2}+\frac{\gamma\left(\gamma+1\right)}{r^2}-\frac{2\alpha}{r}+\varepsilon^2 &0 \\
 0 & -\frac{d^2}{dr^2}+\frac{\gamma\left(\gamma-1\right)}{r^2}-\frac{2\alpha}{r}+\varepsilon^2
\end{pmatrix}\begin{pmatrix}
F\\
G
\end{pmatrix}=0,
\end{equation}
where
\begin{align}\label{alp}
\alpha&=\frac{M}{\rho}\left(\omega{j}-\rho s_1-\frac{\rho s_1s_2}{M}\right),\\\label{varep1}
\varepsilon^2&=k^2-E^2+M^2\omega^2+\left(M+s_2\right)^2.
\end{align}

It can be seen that the equations for $G$ and $F$ have the same mathematical form. Moreover, these equation are related by the change $\gamma\rightarrow\gamma-1$.
Therefore, hereafter we shall focus in the second-order differential equation for the radial function $F$.

\section{The Schr\"odinger factorization method}

In this Section we shall use the Schr\"odinger factorization method \cite{SCH1A,DANIEL} to obtain the energy spectrum and the eigenfunctions of a relativistic quantum particle in cosmic string spacetime.
To accomplish this we first consider the equation (\ref{mard2}) to obtain the following equation
\begin{equation}\label{ecudi2a}
\left(-r^2\frac{d^2}{dr^2}+\varepsilon^2r^2-2\alpha{r}\right)F=-\gamma\left(\gamma+1\right)F.
\end{equation}
In order to construct the $su(1,1)$ algebra generators, we apply the Schr\"odinger
factorization to the left-hand side of equation (\ref{ecudi2a}). Thus, we propose
\begin{equation}\label{sch}
\left(\rho\frac{d}{d\rho}+a\rho+b\right)\left(-\rho\frac{d}{d\rho}+c\rho+f\right)F=gF,
\end{equation}
where $a$, $b$, $c$, $f$ and $g$ are constants to be determined. Expanding this
expression and comparing it with equation (\ref{ecudi2a}) we obtain
\begin{equation}
a=c=\pm\varepsilon,\hspace{2ex}f=1+b=\mp\frac{\alpha}{\varepsilon},\hspace{2ex}
g=b(b+1)-\gamma(\gamma+1).
\end{equation}
Thus, the differential equation (\ref{ecudi2a}) is factorized as
\begin{align}
\left(\mathcal{J}_\mp\mp1\right)\mathcal{J}_\pm F
=\pm\left[\left(\frac{\alpha}{\varepsilon}\pm\frac{1}{2}\right)^2-\left(\gamma\mp\frac{1}{2}\right)^2\right]F,\label{Facdesc1}
\end{align}
where
\begin{equation}
\mathcal{J}_{\pm}=\mp r\frac{d}{dr}+\varepsilon{r}-\frac{\alpha}{\varepsilon}
\end{equation}
are the Schr\"odinger operators. Therefore we can introduce the new pair of operators
\begin{equation}\label{schop}
\mathbb{B}_{\pm}=\mp r\frac{d}{dr}+\varepsilon{r}-{\mathbb{B}}_3,
\end{equation}
with the operator ${\mathbb{B}}_3$ defined as
\begin{equation}\label{tercerop}
{\mathbb{B}}_3F=\frac{1}{2\varepsilon}\left[-r\frac{d^2}{dr^2}+\varepsilon^2r+\frac{\gamma\left(\gamma+1\right)}{r}\right]F=\frac{\alpha}{\varepsilon}F.
\end{equation}

It is easy to show that the operators ${\mathbb{B}}_{\pm}$ and $\mathbb{B}_3$ close the $su(1,1)$
Lie algebra of equation (\ref{com}) of Appendix. Moreover, the quadratic Casimir operator
satisfies the eigenvalue equation
\begin{equation}\label{opcas1}
\mathbb{B}^2F=\gamma(\gamma+1)F.
\end{equation}
The relationship between $\gamma$ and the quantum number $k$ is obtained by using the equation (\ref{cas}) of Appendix. In a similar way we can obtain the relationship between
the group number $n$ and the quantum radial number $n_r$. By comparison of equations (\ref{tercerop}) and (\ref{k0n}) we obtain
\begin{align}
k &=\gamma+1, \quad\quad n=n_r,\\
n+k &=n_r+\gamma+1=\frac{\alpha}{\varepsilon},\label{var}
\end{align}
with $n_r=0,1,2,...$.

The energy spectrum for this problem can be computed from equations (\ref{k0n}), (\ref{varep1}) and (\ref{var}). Thus
\begin{equation}\label{ener1}
E=\left[k^2+M^2\omega^2+\left(M+s_2\right)^2-\frac{M^2\left(\omega j-\rho s_1-\frac{\rho s_1 s_2}{M}\right)^2}{\rho^2\left(n_r+\gamma+1\right)^2}\right]^{1/2}.
\end{equation}

The radial function $F$ can be obtained from an analytical approach. If we consider the asymptotic behavior of the solutions of the radial differential
equation (\ref{ecudi2a}) for $r\rightarrow 0$ and $r\rightarrow \infty$ we propose
\begin{equation}\label{fpro}
F=r^{\gamma+1}e^{-\varepsilon{r}}f(r).
\end{equation}
Hence, if we introduce the new variable $y =2\varepsilon{r}$, the function $f(r)$ must satisfy
\begin{equation}
\left[y^2\frac{d^2}{dy^2}+2y(\gamma-\frac{y}{2}+1)\frac{d}{dy}+\left(\frac{M\left(\omega j-\rho s_1-\frac{\rho s_1 s_2}{M}\right)}{\rho\varepsilon}-\gamma-1\right)y\right]f\left(\frac{y}{2\varepsilon}\right)=0,
\end{equation}
The solution for this equation is the confluent hypergeometric function. Thus from equation (\ref{fpro}) we obtain
\begin{equation}\label{fsol}
F=\mathcal{C}_1r^{\gamma+1}e^{-\varepsilon{r}}\;_1F_1\left(-n+1,2\gamma+2;2\varepsilon{r}\right).
\end{equation}
The other radial function $G$ can be easily obtained by performing the change $\gamma\rightarrow\gamma-1$. However, it is important to note that, since both functions belong to the same energy
level, the energy spectrum of equation (\ref{ener1}) imposes the change $n\rightarrow n+1$. Hence,
\begin{equation}\label{gsol}
G=\mathcal{C}_2r^{\gamma}e^{-\varepsilon{r}}\;_1F_1\left(-n,2\gamma;2\varepsilon{r}\right).
\end{equation}
These functions can be written in terms of the Laguerre polynomials by using the relationship
\begin{equation}
_1F_1\left(-n,m+1;x\right)=\frac{m!n!}{(m+n)!}L_n^m(x).
\end{equation}
Therefore, the radial functions $F$ and $G$ can be expressed by the following spinor
\begin{equation}
\Phi=\begin{pmatrix}
F\\
G
\end{pmatrix}=\begin{pmatrix}
\mathcal{C}_1r^{\gamma+1}e^{-\varepsilon{r}}L_{n-1}^{2\gamma+1}\left(2\varepsilon r\right)\\
\mathcal{C}_2r^{\gamma}e^{-\varepsilon{r}}L_{n}^{2\gamma-1}\left(2\varepsilon r\right)
\end{pmatrix}.
\end{equation}

The ground state of the spinor $\Phi$ is obtained for $n=0$. However, since the Laguerre polynomials $L_n^{(\alpha)}$ are not defined if $n$ is a negative integer, we conclude that,
up to a normalization constant
\begin{equation}\label{estsch}
\Phi_{0,Sch}=\begin{pmatrix}
0\\
r^\gamma e^{-\frac{M}{\rho\gamma}\left(\omega j-\rho s_1-\frac{\rho s_1 s_2}{M}\right)r}
\end{pmatrix}.
\end{equation}
On the other hand, the SUSY operators can be defined from equation (\ref{mard1}) as
\begin{equation}
\mathcal{A}^{\pm}=\pm\frac{d}{dr}+\frac{\gamma}{r}-\frac{M}{\rho\gamma}\left(\omega j-\rho s_1-\frac{\rho s_1 s_2}{M}\right).
\end{equation}
From these operators we obtain that the partner Hamiltonians-type are given by
\begin{equation}
H_+=\mathcal{A}^-\mathcal{A}^+,\hspace{1.0cm} H_-=\mathcal{A}^+\mathcal{A}^-.
\end{equation}
The ground state for the Hamiltonian $H_-$ satisfies the condition $A^-\Phi_{0,SUSY}=0$, which leads to
\begin{equation}
\Phi_{0,SUSY}=r^\gamma e^{-\frac{M}{\rho\gamma}\left(\omega j-\rho s_1-\frac{\rho s_1 s_2}{M}\right)r}.
\end{equation}
In a similar way, the solution for the equation $A^+\Phi_{0,SUSY}=0$ is
\begin{equation}
\Phi_{0,SUSY}=r^{-\gamma} e^{\frac{M}{\rho\gamma}\left(\omega j-\rho s_1-\frac{\rho s_1 s_2}{M}\right)r}.
\end{equation}
Notice that this function is not square-integrable, and therefore, is not a physically acceptable solution. Thus, must be taken as the zero function. Hence, the SUSY ground-state is given by
\begin{equation}\label{estsusy}
\Phi_{0,SUSY}=
\begin{pmatrix}
0\\
r^\gamma e^{-\frac{M}{\rho\gamma}\left(\omega j-\rho s_1-\frac{\rho s_1 s_2}{M}\right)r}
\end{pmatrix}.
\end{equation}
From equations (\ref{estsusy}) and (\ref{estsch}) we conclude that the Schr\"odinger and
SUSY ground states are the same.

The radial functions $F_+$ and $G_-$ for the relativistic quantum particle in cosmic string spacetime can be obtained from equation (\ref{minver}) and explicitly are
\begin{equation}\label{spinor}
\begin{pmatrix}
F_+\\
G_-
\end{pmatrix}= \mathbb{M}\begin{pmatrix}
F\\
G
\end{pmatrix}= r^{\gamma}e^{-\varepsilon{r}}
\begin{pmatrix}
\mathcal{C}_1\left(\gamma-\frac{j}{\rho}\right){r}L_{n-1}^{2\gamma+1}(2\varepsilon{r})-\mathcal{C}_2s_1 L_{n}^{2\gamma-1}(2\varepsilon{r})\\
\mathcal{C}_1s_1{r}L_{n-1}^{2\gamma+1}(2\varepsilon{r})+\mathcal{C}_2\left(\gamma-\frac{j}{\rho}\right)L_{n}^{2\gamma-1}(2\varepsilon{r})
\end{pmatrix}.
\end{equation}

Therefore, we have used the Schr\"odinger factorization method and the theory of unitary representation to obtain the energy spectrum
and the eigenfunction of a relativistic quantum particle in cosmic string spacetime. Moreover, we showed that the ground state of
this problem coincides with the ground state of the SUSY quantum theory.

In references\cite{MSAL1,MSAL2} we have studied the relativistic Kepler-Coulomb problem and the Dirac equation with Coulomb-type scalar and vector potential in $D+1$ dimensions from an algebraic approach by using the Schr\"odinger factorization to construct the $su(1,1)$ algebra generators.

\section{The tilting transformation method}

In this Section we shall study this problem by using an alternative approach, the tilting transformation. Thus, we can introduce the operators \cite{AOB,KTH}
\begin{align}\label{A1}
A_0&=\frac{1}{2}\left(rP_r^2+\frac{\gamma(\gamma+1)}{r}+r\right),\\\label{A2}
A_1&=\frac{1}{2}\left(rP_r^2+\frac{\gamma(\gamma+1)}{r}-r\right),\\\label{A3}
A_2&=rP_r=-ir\left(\frac{d}{dr}+\frac{1}{r}\right),
\end{align}
where the operator $P_r^2$ is defined as
\begin{equation}
P_r^2=-\frac{d^2}{dr^2}-\frac{2}{r}\frac{d}{dr}.
\end{equation}
These new operators also close the $su(1,1)$ Lie algebra. The difference between the operators obtained from the Schr\"odinger factorization and these operators is that the last ones do not depend on the rescaling radial variable (which includes the energy).

In the same way it was done for the Schr\"odinger factorization, we shall focus just on the radial equation for the eigenfunction $F$ of equation (\ref{mard2}). If we define $F(r)=r\widetilde{F}(r)$, equation (\ref{mard2}) becomes
\begin{equation}\label{estfi}
\widetilde{\Delta}(E)|\widetilde{F}(r)\rangle=0,
\end{equation}
where
\begin{equation}
\widetilde{\Delta}(E)\equiv\left[-\frac{d^2}{dr^2}-\frac{2}{r}\frac{d}{dr}+\frac{\gamma\left(\gamma+1\right)}{r^2}-\frac{2\alpha}{r}+\varepsilon^2\right].
\end{equation}
In this equation $|\widetilde{F}(r)\rangle$ represents the physical states and $\alpha$ and $\varepsilon^2$ are defined in equations (\ref{alp}) and (\ref{varep1}).
We can use the operators $A_0$, $A_1$ and $A_2$ to write the equation (\ref{estfi}) as follows
\begin{equation}\label{xibar}
\left[\frac{1}{2}\left(A_0+A_1\right)+\frac{1}{2}\left(\varepsilon^2\right)\left(A_0-A_1\right)-\alpha\right]\widetilde{F}(r)=0.
\end{equation}

By introducing the scaling operator $e^{i\theta A_2}$ and using the Baker-Campbell-Hausdorff formula it is easy to show the formulas
\begin{align}
e^{-i\theta A_2}A_0e^{+i\theta A_2}&=A_0\cosh(\theta)+A_1\sinh(\theta),\\
e^{-i\theta A_2}A_1e^{+i\theta A_2}&=A_0\sinh(\theta)+A_1\cosh(\theta).
\end{align}
From these equations it follows that
\begin{equation}
e^{-i\theta A_2}(A_0\pm A_1)e^{i\theta A_2}=e^{\pm\theta}(A_0\pm A_1).
\end{equation}
Thus, from above expression the equation (\ref{xibar}) can be written as
\begin{equation}\label{delt1}
\Delta(E)|\overline{F}(r)\rangle=\left[\frac{1}{2}\left(e^{\theta}+\varepsilon^2e^{-\theta}\right)A_0+\frac{1}{2}\left(e^{\theta}-\varepsilon^2e^{-\theta}\right)A_1-\alpha\right]|\overline{F}(r)\rangle=0,
\end{equation}
where $|\overline{F}(r)\rangle=e^{-i\theta A_2}|\widetilde{F}(r)\rangle$ and $\Delta(E)=e^{-i\theta A_2}\widetilde{\Delta}(E)e^{i\theta A_2}$. The scaling parameter $\theta$ can be chosen such that the coefficient of $A_1$ vanishes. With $\theta=\ln\varepsilon$ we obtain
\begin{equation}\label{ecde}
\Delta(E)|\overline{F}(r)\rangle=\left(\varepsilon{A_0}-\alpha\right)|\overline{F}(r)\rangle=0.
\end{equation}
If we take $|\overline{F}(r)\rangle$ as an $SU(1,1)$ group state $|\overline{n},\overline{k}\rangle$, the energy spectrum can be obtained from this relationship and equation (\ref{k0n}) of Appendix. By remembering the relationship between the physical quantum numbers $n_r, \gamma$ and the group numbers $n, k$ we obtain
\begin{equation}\label{enecoe}
E=\left[k^2+M^2\omega^2+\left(M+s_2\right)^2-\frac{M^2\left(\omega j-\rho s_1-\frac{\rho s_1 s_2}{M}\right)^2}{\rho^2\left(n_r+\gamma+1\right)^2}\right]^{1/2}.
\end{equation}
This energy spectrum matches with that obtained in equation (\ref{ener1}) by using Schr\"odinger factorization method. Moreover, this also coincides with the energy spectrum obtained in reference \cite{YESI}.

The eigenfunctions basis for the irreducible unitary representation of the $su(1,1)$ Lie algebra (Sturmian basis) is \cite{MOS,CCG}
\begin{equation}
\langle r|\overline{n},\overline{\gamma}\rangle=\overline{F}_{n\gamma}(r)=2\sqrt{\frac{\Gamma\left(n\right)}{\Gamma\left(n+2\gamma+1\right)}}\left(2r\right)^\gamma e^{-r}L_{n-1}^{2\gamma+1}\left(2r\right),
\end{equation}
and the lower component is obtained by performing the change $\gamma\rightarrow\gamma-1$, which implies a change $n\rightarrow n+1$, since both eigenfunctions belong to the same energy level. Hence
\begin{equation}
\overline{G}_{n\gamma}(r)=2\sqrt{\frac{\Gamma\left(n+1\right)}{\Gamma\left(n+2\gamma\right)}}\left(2r\right)^{\gamma-1} e^{-r}L_{n}^{2\gamma-1}\left(2r\right).
\end{equation}
By using the relation
\begin{equation}\label{scall}
e^{i\theta{A_2}}f(r)=e^{\theta}f(e^{\theta}r)
\end{equation}
where $f(r)$ is an arbitrary spherically symmetric function, the physical states can be written as
\begin{align}
\widetilde{F}(r)&=A_n(2\varepsilon{r})^\gamma e^{-\varepsilon{r}}L_{n-1}^{2\gamma+1}(2\varepsilon{r}),\\
\widetilde{G}(r)&=B_n(2\varepsilon{r})^{\gamma-1} e^{-\varepsilon{r}}L_{n}^{2\gamma-1}(2\varepsilon{r}).
\end{align}
Therefore, remembering that $F=r\widetilde{F}(r)$ and $G=r\widetilde{G}(r)$ we obtain
\begin{align}
F(r)&=A_n(2\varepsilon)^\gamma r^{\gamma+1}e^{-\varepsilon{r}}L_{n-1}^{2\gamma+1}(2\varepsilon{r}),\\
G(r)&=B_n(2\varepsilon)^{\gamma-1} r^{\gamma}e^{-\varepsilon{r}}L_{n}^{2\gamma-1}(2\varepsilon{r}).
\end{align}

From equation (\ref{minver}) we obtain the radial functions $F_+(r)$ and $G_-(r)$ in terms of the group states $\overline{F}(r)$ and $\overline{G}(r)$
\begin{equation}\label{funrad}
\begin{pmatrix}
F_+\\
G_-
\end{pmatrix}= (2\varepsilon)^{\gamma-1}r^{\gamma}e^{-\varepsilon{r}}
\begin{pmatrix}
A_n(2\varepsilon)\left(\gamma-\frac{j}{\rho}\right){r}L_{n-1}^{2\gamma+1}(2\varepsilon{r})-B_ns_1 L_{n}^{2\gamma-1}(2\varepsilon{r})\\
A_n(2\varepsilon)s_1{r}L_{n-1}^{2\gamma+1}(2\varepsilon{r})+B_n\left(\gamma-\frac{j}{\rho}\right)L_{n}^{2\gamma-1}(2\varepsilon{r})
\end{pmatrix}.
\end{equation}

Therefore, by comparing this expression with equation (\ref{mard2}) we can conclude that the eigenfunctions $F_+$ and $G_-$ obtained from both methods, the Schr\"odinger factorization and the tilting transformation are the same. The relationship between the normalization coefficients $A_n$ and $B_n$ can be obtained by evaluating the first coupled equation of (\ref{mard1}) in the limit $r\rightarrow0$. By using the formula
\begin{equation}
L_n^{\lambda}(0)=\frac{\Gamma(n+\lambda+1)}{n!\Gamma(\lambda+1)},
\end{equation}
the relationship between $A_n$ and $B_n$ is
\begin{equation}
B_n=\frac{\left(n+2\gamma\right)n\varepsilon}{\gamma{\eta}}A_n,
\end{equation}
where $\eta=\sqrt{E^2-k^2}+\frac{M\omega\rho{s_1}+(M+s_2)j}{\gamma\rho}$.

Thus, the radial spinor for the  the relativistic quantum particle in cosmic string spacetime can be explicitly written as
\begin{equation}\label{funexp}
\begin{pmatrix}
F_+\\
G_-
\end{pmatrix}= A_n(2\varepsilon)^{\gamma}r^{\gamma}e^{-\varepsilon{r}}
\begin{pmatrix}
R_1{r}L_{n-1}^{2\gamma+1}(2\varepsilon{r})-R_2L_{n}^{2\gamma-1}(2\varepsilon{r})\\
T_1{r}L_{n-1}^{2\gamma+1}(2\varepsilon{r})+T_2L_{n}^{2\gamma-1}(2\varepsilon{r})
\end{pmatrix},
\end{equation}
where
\begin{equation}
R_1=\frac{\gamma\rho-j}{\rho},\hspace{0.3cm} R_2=\frac{n(n+2\gamma){s_1}}{2\gamma\rho\eta},\hspace{0.3cm}
T_1=s_1,\hspace{0.3cm} T_2=\frac{n(n+2\gamma)(\gamma\rho-j)}{2\gamma\rho\eta}.
\end{equation}

The normalization coefficient $A_n$ can be computed by means of the relativistic normalization. The discrete symmetry $G_{+}=\lambda{F_{+}}$ and $F_{-}=\lambda{G_{-}}$
let us write the normalization as follows
\begin{equation}\label{normre}
\int_0^{\infty}\left(1+\lambda^2\right)r\left(F_+^*F_++G_-^*G_-\right)dr=1.
\end{equation}

These integrals can be calculated by using the following Laguerre integrals
\begin{equation}
\int_0^{\infty}e^{-x}x^{\alpha+2}\left[L_{n-1}^{\alpha}(x)\right]^2dx=\frac{\Gamma(n+\alpha)}{\Gamma(n)}\left[6(n-1)(n+\alpha)+(\alpha+1)(\alpha+2)\right],
\end{equation}
\begin{equation}
\int_0^{\infty}e^{x}x^{\alpha+1}L_{n-1}^{\alpha}(x)L_{n}^{\alpha-2}(x)dx=-\frac{\Gamma(n+\alpha+1)}{\Gamma(n)}\left(5n+3\alpha-3\right).
\end{equation}
By using these results the coefficient $A_n$ is
\begin{equation}
A_n=2\varepsilon^2\left[\frac{\rho{(n-1)!}}{\left(1+\lambda^2\right)\gamma(\gamma\rho-j)\Gamma(n+2\gamma+1)\left[\Theta+\varepsilon^2\sigma(\Theta+2\gamma)\right]}\right]^{\frac{1}{2}},
\end{equation}
where we have introduced the variables
\begin{equation}
\Theta=3n^2+\gamma(6n+2\gamma-1), \hspace{0.3cm} \sigma=\frac{n(n+2\gamma)}{(\gamma\eta)^2}.
\end{equation}

\section{$SU(1,1)$ relativistic coherent states}

In this Section we shall construct the relativistic coherent states for the radial functions $F_+$ and $G_-$ by using the Sturmian basis and the transformation given in equation (\ref{funrad}).
The $SU(1,1)$ Perelomov coherent states are defined as \cite{PER}
\begin{equation}
|\zeta\rangle=D(\xi)|k,0\rangle=\left(1-|\xi|^2\right)^k\sum_{s=0}^\infty\sqrt{\frac{\Gamma(n+2k)}{s!\Gamma(2k)}}\xi^s|k,0\rangle,
\end{equation}
with $D(\xi)$ the displacement operator and $|k,0\rangle$ the lowest normalized state.
Thus, we can apply the operator $D(\xi)$ to the ground states of the functions $\overline{F}$ and $\overline{G}$
\begin{align}\label{est1}
\overline{F}(r,\xi)&=\frac{2\left(1-|\xi|^2\right)^{\gamma+1}}{\sqrt{\Gamma(2\gamma+2)}}(2r)^\gamma e^{-r}\sum_{n=1}^\infty\xi^{n-1}L_{n-1}^{2\gamma+1}(2r),\\\label{est2}
\overline{G}(r,\xi)&=\frac{2\left(1-|\xi|^2\right)^{\gamma}}{\sqrt{\Gamma(2\gamma)}}(2r)^{\gamma-1} e^{-r}\sum_{n=0}^\infty\xi^{n}L_{n}^{2\gamma-1}(2r).
\end{align}
This sum can be computed from the Laguerre polynomials generating function
\begin{equation}
\sum_{n=0}^\infty L_n^\nu(x)=\frac{e^{-xy/(1-y)}}{(1-y)^{\nu+1}}.
\end{equation}

The physical coherent states $F(r,\xi)$ and $G(r,\xi)$ can be constructed from equations (\ref{est1}) and (\ref{est2}). By using equation (\ref{scall}) and multiplying by $r$ we obtain
\begin{align}
F(r,\xi)&=C_n\frac{2\left(1-|\xi|^2\right)^{\gamma+1}}{\sqrt{\Gamma(2\gamma+2)}(1-\xi)^{2\gamma+2}}(2\varepsilon)^\gamma r^{\gamma+1} e^{\frac{-\varepsilon{r}\left(1+\xi\right)}{1-\xi}},\\
G(r,\xi)&=D_n\frac{2\left(1-|\xi|^2\right)^{\gamma}}{\sqrt{\Gamma(2\gamma)}(1-\xi)^{2\gamma}}(2\varepsilon)^{\gamma-1} r^{\gamma} e^{\frac{-\varepsilon{r}\left(1+\xi\right)}{1-\xi}}.
\end{align}
The relationship between the normalization coefficients $C_n$ and $D_n$ can be find in the same way as was done for $A_n$ and $B_n$. Since $F(r,\xi)$ and $G(r,\xi)$ also satisfy the coupled equations (\ref{mard1}), from the first coupled equation of (\ref{mard1}) we obtain in the limit $r\rightarrow 0$ that
\begin{equation}
D_n=2\frac{\varepsilon\left(1-|\xi|^2\right)(2\gamma+1)}{\left(1-\xi\right)^2\eta}\sqrt{\frac{\Gamma(2\gamma)}{\Gamma(2\gamma+2}}C_n.
\end{equation}
Therefore, the $SU(1,1)$ radial coherent states for the relativistic quantum particle in cosmic string spacetime can be expressed in a matricial form as
\begin{equation}\label{funexpco}
\begin{pmatrix}
F_+\\
G_-
\end{pmatrix}= \frac{2C_n\left(1-|\xi|^2\right)^{\gamma+1}}{\left(1-\xi\right)^{2\gamma+2}\sqrt{\Gamma(2\gamma+2)}}(2\varepsilon)^\gamma r^\gamma e^{\frac{-\varepsilon{r}(1+\xi)}{(1-\xi)}}
\begin{pmatrix}
\frac{(\gamma\rho-j)r}{\rho}-\frac{s_1(2\gamma+1)}{\eta}\\
s_1{r}+\frac{(2\gamma+1)(\gamma\rho-j)}{\eta\rho}
\end{pmatrix}.
\end{equation}
In order to obtain the normalization coefficient $C_n$ we can use the relativistic normalization given in equation (\ref{normre}). The integrals that arise are computed in terms of gamma functions. Thus, $C_n$ explicitly is
\begin{equation}
C_n=\left[\frac{\rho{\varepsilon^2}\left(1-\xi^2\right)^{2\gamma}}{\left(1+\lambda^2\right)2\gamma(\gamma\rho-j)\left(1-|\xi|^2\right)^{2\gamma+2}\left(\sigma^{'}+\Theta^{'}\right)}\right]^{\frac{1}{2}},
\end{equation}
where
\begin{equation}
\sigma^{'}=\frac{\left(2\gamma+1\right)^2}{\eta^2\left(1-\xi^2\right)^2},\hspace{0.5cm} \Theta^{'}=\frac{\Gamma(2\gamma+4)}{(2\epsilon)^2\Gamma(2\gamma+2)\left(1+\xi\right)^4}.
\end{equation}
Therefore, we have constructed the relativistic coherent states for the radial part of a quantum particle in cosmic string spacetime.

In reference \cite{PLA} we have constructed the relativistic coherent states for the Dirac-Kepler-Coulomb problem in $D+1$ dimensions with scalar and vector potentials by using the tilting transformation method.
In an analogous way, in reference \cite{CHN} we studied the problem of a Dirac-Moshinsky oscillator coupled to an external magnetic field by using the an appropriate $su(1,1)$ Lie algebra generators. This symmetry
let us compute the relativistic coherent states for this problem.

\renewcommand{\theequation}{A.\arabic{equation}}
\setcounter{equation}{0}
\section*{Appendix: The $SU(1,1)$ Group and its coherent states}

The $su(1,1)$ Lie algebra is spanned by the generators $K_{+}$, $K_{-}$
and $K_{0}$, which satisfy the commutation relations \cite{VOU}
\begin{eqnarray}
[K_{0},K_{\pm}]=\pm K_{\pm},\quad\quad [K_{-},K_{+}]=2K_{0}.\label{com}
\end{eqnarray}
The action of these operators on the Fock space states
$\{|k,n\rangle, n=0,1,2,...\}$ is
\begin{equation}
K_{+}|k,n\rangle=\sqrt{(n+1)(2k+n)}|k,n+1\rangle,\label{k+n}
\end{equation}
\begin{equation}
K_{-}|k,n\rangle=\sqrt{n(2k+n-1)}|k,n-1\rangle,\label{k-n}
\end{equation}
\begin{equation}
K_{0}|k,n\rangle=(k+n)|k,n\rangle,\label{k0n}
\end{equation}
where $|k,0\rangle$ is the lowest normalized state. The Casimir
operator for any irreducible representation satisfies
\begin{equation}
K^{2}=-K_{+}K_{-}+K_{0}(K_{0}-1)=k(k-1).\label{cas}
\end{equation}
The theory of unitary irreducible representations of the $su(1,1)$ Lie algebra has been
studied in several works \cite{ADAMS} and it is based on equations (\ref{k+n})-(\ref{cas}). 
Thus, a representation of $su(1,1)$ algebra is determined by the number $k$. For the purpose of the
present work we will restrict to the discrete series only, for which
$k>0$.

The $SU(1,1)$ Perelomov coherent states $|\zeta\rangle$ are
defined as \cite{PER}
\begin{equation}
|\zeta\rangle=D(\xi)|k,0\rangle,\label{defPCS}
\end{equation}
where $D(\xi)=\exp(\xi K_{+}-\xi^{*}K_{-})$ is the displacement
operator and $\xi$ is a complex number. From the properties
$K^{\dag}_{+}=K_{-}$ and $K^{\dag}_{-}=K_{+}$ it can be shown that
the displacement operator possesses the property
\begin{equation}
D^{\dag}(\xi)=\exp(\xi^{*} K_{-}-\xi K_{+})=D(-\xi),
\end{equation}
and the so called normal form of the displacement operator is given by
\begin{equation}
D(\xi)=\exp(\zeta K_{+})\exp(\eta K_{0})\exp(-\zeta^*
K_{-})\label{normal},
\end{equation}
where $\xi=-\frac{1}{2}\tau e^{-i\varphi}$, $\zeta=-\tanh
(\frac{1}{2}\tau)e^{-i\varphi}$ and $\eta=-2\ln \cosh
|\xi|=\ln(1-|\zeta|^2)$ \cite{GER}. By using this normal form of the displacement
operator and equations (\ref{k+n})-(\ref{k0n}), the Perelomov coherent states are found to
be \cite{PER}
\begin{equation}
|\zeta\rangle=(1-|\xi|^2)^k\sum_{s=0}^\infty\sqrt{\frac{\Gamma(n+2k)}{s!\Gamma(2k)}}\xi^s|k,s\rangle.\label{PCN}
\end{equation}

\section{Concluding remarks}

We obtained the energy spectrum and the eigenfunctions for a relativistic quantum particle in cosmic string spacetime in the presence
of a uniform magnetic field and scalar potential. In our analysis, we supposed that the magnetic field was parallel to
the string. The uncoupled second order radial equations were obtained by using a transformation matrix which diagonalized the
$1/r$ term. It is shown that these two uncoupled equations are related by the change $\gamma\rightarrow\gamma-1$.

In order to solve our problem we considered one of the two uncoupled second order radial equations. We obtained the energy spectrum
and the eigenfunctions by using the theory of unitary representations in two different algebraic methods. In the first one, we applied
the Schr\"odinger factorization and in the second one we used the tilting transformation.

The problem we have treated in the present paper has been previously studied in references \cite{MAD} and \cite{YESI}. In these works, the uncoupled equations for the spinor
components were obtained by using a different diagonalization matrix. More specifically, in reference \cite{MAD} the authors
considered the cases of relativistic spin-$0$ and spin-$\frac{1}{2}$ particles, where the energy spectrum and eigenfunctions are found in an analytical way.
On the other hand, in reference \cite{YESI}, the author solved this problem in an algebraic way by using the Schr\"odinger factorization method.
However, the parameters used in this factorization are different to those we used in the present work.

From the Sturmian basis of the $su(1,1)$ Lie algebra we constructed the Perelomov coherent states for each radial component. The
complete relativistic coherent states for a relativistic quantum particle in cosmic string spacetime in the presence
of a uniform magnetic field and scalar potential is obtained by using the transformation matrix previously introduced in Section $2$.

\section*{Acknowledgments}
This work was partially supported by SNI-M\'exico, COFAA-IPN,
EDI-IPN, EDD-IPN, SIP-IPN project number $20161727$.

\end{document}